\documentstyle[openbib,12pt]{article}
\begin{document}
\def \blank{\mbox{}}

\def\A{\mbox{\bf A}}
\def\B{\mbox{\bf B}}
\def\C{\mbox{\bf C}}
\def\M{\mbox{\bf M}}
\def\P{\mbox{\bf P}}
\def\x{\mbox{{\bf x}}}
\def\Q{\mbox{{\bf Q}}}
\def\R{\mbox{{\bf R}}}
\def\S{\mbox{{\bf S}}}
\def\L{\mbox{$L$}}
\def\T{\mbox{$T$}}
\def\Ec{\mbox{{$\cal E$}}}
\def\Pc{\mbox{{$\cal P$}}}
\def\Ac{\mbox{{$\cal A$}}}
\def\eps{\mbox{$\epsilon$}}
\def\Y{\mbox{$\bf{Y}$}}
\def\y{\mbox{$\bf{y}$}}
\def\v{\mbox{$\bf{y}$}}
\def\z{\mbox{$\bf{z}$}}
\def\e{\mbox{$\epsilon$}}
\def\f{\mbox{$\bf{f}$}}
\def\F{\mbox{$\bf{F}$}}
\def\E{\mbox{$\bf{E}$}}
\def\H{\mbox{$\bf{H}$}}
\def\G{\mbox{$\bf{G}$}}
\def\U{\mbox{$\bf{U}$}}
\def\J{\mbox{$\bf{J}$}}
\def\doublespace{\parskip 4pt plus 1.5pt
                \baselineskip 24pt plus 1pt minus .5pt
                \lineskip 6pt plus 1pt \lineskiplimit 5pt}
\newcommand{\be}{\begin{equation}}
\newcommand{\bea}{\begin{eqnarray}}
\newcommand{\ee}{\end{equation}}
\newcommand{\eea}{\end{eqnarray}}

\begin{titlepage}

\begin{center}
{\Large{\bf {Sensory Coding with Dynamically Competitive
Networks}}}
\bigskip
\bigskip
\bigskip
\bigskip

M. I. Rabinovich,\\
rabin@poincare.ucsd.edu\\
\bigskip
R. Huerta,\\
huerta@routh.ucsd.edu\\
\bigskip
A. Volkovskii,\\
volkovsk@routh.ucsd.edu\\
\bigskip
Institute for Nonlinear Science,\\
\bigskip
\bigskip
Henry D. I. Abarbanel\\
hdia@jacobi.ucsd.edu\\
Marine Physical Laboratory\\
Scripps Institution of Oceanography,\\
Department of Physics\\

\bigskip
\bigskip
University of California, San Diego\\
La Jolla, CA 93093-0402,\\
\bigskip
and\\
G. Laurent,\\
laurentg@cco.caltech.edu\\
Division of Biology\\
California Institute of Technology\\
Pasadena, CA 91125\\

\bigskip
\,\,{\today}
\bigskip

\end{center}
\end{titlepage}

\begin{abstract}

Studies of insect olfactory processing indicate that odors are
represented by rich {\bf spatio-temporal} patterns of neural
activity.
These patterns are very difficult to predict {\it a priori}, yet they
are  stimulus specific and reliable upon repeated stimulation with
the
same input. We formulate here a theoretical framework in which
we can
interpret these experimental results. We propose a paradigm of
``dynamic competition'' in which inputs (odors) are represented
by
internally competing neural assemblies. Each pattern is the result
of
dynamical motion within the network and does not involve a
``winner''
among competing possibilities. The model produces
spatio-temporal
patterns with strong resemblance to those observed
experimentally and
possesses many of the general features one desires for pattern
classifiers: large information capacity, reliability, specific
responses to specific inputs, and reduced sensitivity to initial
conditions or influence of noise. This form of neural processing
may
thus describe the organizational principles of neural information
processing in sensory systems and go well beyond the
observations on
insect olfactory processing which motivated its development.
\end{abstract}

Sensory networks in the brain effortlessly solve very complex
pattern
learning, pattern storage, and pattern recognition problems.
These
problems are complex because the natural environments in
which animals
live are noisy and nonstationary and because natural patterns
such as
faces, animal calls or odors are multidimensional, unpredictable
by
the animal, and remarkably numerous. This ease of computation
and
striking reliability is perhaps perplexing as the computing
elements
of the brain, our neurons, are often considered to be unreliable
electrical devices. Just how the brain represents, stores, and
recalls
complex sensory inputs using these neurons remains a major
challenge
in neuroscience.

Many agree that the brain uses distributed and combinatorial
codes for
sensory representations: codes in which input-specific
information is
contained across populations of coarsely tuned neurons
\cite{A94},\cite{G95}. However, the possibility that the dynamics
of
neural circuits, that is the temporal behavior exhibited by groups
of
interconnected neurons\cite{FREK97}, \cite{GKES89},
\cite{SG95},
\cite{ABMV93}, \cite{VHAB95}, \cite{PVBH98} are also centrally
important in sensory coding is not yet widely accepted. There
appear
to be two reasons for this reluctance. First, while stimulus or
behavior-specific temporal patterns of neural activity have been
observed in a few systems, their relevance for perception or
action
has so far been difficult to establish directly. Second, in the few
cases where information theory has been applied to the analysis
of
single neuron responses, the temporal features of those
responses
accounted for only a small fraction of their variance
\cite{TRSC96},
\cite{ROS90}, \cite{RO90}, \cite{MORG91}, \cite{VP96}. As
applied to
single neurons, then, average firing rate measurements
appeared to
convey most of the information that could be theoretically
extracted
from sensory spike trains\footnote{We do not consider here the
case in
which the timing of spikes represents the time variations of the
stimulus. Rather, we focus on neural dynamics as an intrinsic
component of a stimulus representation and not on its time
evolution
alone}.

\section*{Dynamical Odor Representation: Experimental Data}

Experiments conducted in our laboratory and elsewhere directly
address
these two objections. The observations show that (1) temporal
and
relational features of neural activity contain information which is
not captured in firing rates and cannot be found by considering
single
neurons alone \cite{LWD96}, \cite{WL96} and (2) such
temporal/relational features are required for certain sensory
discrimination tasks \cite{SBSL97} as well as for optimal
decoding by
downstream neurons \cite{MBL98}. Circuit dynamics thus appear
relevant
not only for an observer deciphering brain codes but also, and
more
importantly, for the animal's own sensory performance.

What form do these temporal patterns take? Our observations
focused on
olfactory processing in locust and bees. The insect antennal
lobes,
the analog of the vertebrate olfactory bulb \cite{HS97}, contain
two
classes of neurons : (1) the excitatory projection neurons (PN),
the
analog of mitral cells in vertebrates, which send signals
downstream
to other brain structures, and (2) inhibitory local neurons (LN),
the
analog to granule cells in vertebrates, whose projections are
within
the antennal lobe only. In locust all odor related information is
carried by 830 PNs from each antennal lobe \cite{LL96}. Each
odor,
regardless of its complexity, evokes synchronized and oscillatory
activity in the 20-30 Hz range, but only in a specific subset of
these
PNs \cite{LD94}. Probably no more than 20\% are excited by any
particular odor. Individual PNs, however, display slower odor and
PN-specific response patterns superimposed on their fast global
oscillatory coordination. Consequently, pairwise synchronization
of
PNs is usually transient \cite{LWD96}, \cite{WL96}. An odor
representation can thus be thought of as an evolving PN {\it
assembly}
whose members are progressively updated throughout the
duration of a
stimulus. Although PN update is orchestrated at an overall rate
of
20-30 Hz, the slower aspects of individual PN responses do not
appear
to contain periodic motifs.

If an odor stimulus is sustained for several seconds, olfactory
receptor adaptation diminishes the PN/LN network activation. We
can thus experimentally study the spatio-temporal evolution of
this system for a limited time during each presentation of the
stimulus. Nonetheless, we observe that this evolution remains
stable over repeated, identical stimulations. This leads us to the
central question of this
paper: What advantages, if any, could this rich, distributed
spatio-temporal behavior present for stimulus representation or
classification?

To answer this question, we must first understand the origin of
the
dynamical aspects of this representation. Only then will we be
able to
construct models which embody these dynamical rules. Two of
the main
experimental features we want to capture are (1)
stimulus-evoked
nonstationary and irregular spatio-temporal waveforms, or
patterns,
and (2) stability of these patterns in time and neuron space over
repeated presentations of the stimulus, despite the natural
fluctuations in the stimuli and in the neural system. To achieve
these
observed biological features, we cannot use the now familiar
strategy
for representing odors and other sensory inputs which seeks to
associate a basin of attraction - more particularly the attractor in
that basin - with each specific input. We have developed a
different
strategy.

\section*{Dynamical Competition: Principles }

We hypothesize, guided by our experiments, that the transient
dynamics
of the system embodied in its state space trajectory is dictated
by
the input itself. The state space of the autonomous
(nonstimulated)
system  is enlarged by the degrees of freedom contained in the
stimulus upon reception of the input, and dynamical evolution
within
that space occurs in an input dependent way. The trajectory
followed
by the system then defines the only global attractor associated
with
the particular input. In other words the system does not seek
attractors in pre-existing basins of attraction defined by the
network
alone.

We describe below a set of simplified, biologically realistic
mathematical models which realize this idea. They produce a
dynamical
evolution which does not go to an attractor of the stimulus free
system but rather realizes a rich set of input-dependent
trajectories.

The basic principle relies on a dynamical competition between
evolving groups of PNs interconnected by inhibition through the
LNs. In state space this continuing competition is embodied in
trajectories in the neighborhood of heteroclinic orbits, appearing
as a sequence of ``ribs'' in state space,
connecting unstable fixed points or limit cycles of the system.
In this view of sensory information processing, the solution to
the decoding/recognition problem lies not in the attractors of the
autonomous dynamical systems but in the orbits themselves. A
state space
representation of this kind of trajectory is displayed in Figure 1a.

As observed in experiments, specific odors are represented not
only by specific groups of excited neurons  but also by the
temporal sequence of their activation.
To reproduce the very specific
stimulus dependent transient
behaviors seen in biology the system must be strongly
dissipative. Under such conditions,  a trajectory, initiated by a
given
input, rapidly converges to the heteroclinic  sequence
associated with the particular stimulus. This rapid convergence
is essential for the reliability of this mode of representation.

There are surely many possible implementations of this
general
idea. We provide here examples which are appealing because of
their biological underpinnings, their relative simplicity and their
potential generality.

\section*{Dynamical Models: Architecture}

The main features of our model are

\begin{itemize}
\item The network can be described as a set of neural
ensembles or layers coupled by inhibition and excitation.

\item Each ensemble contains two types of neurons: excitatory
PNs and inhibitory LNs. In our model the PNs and LNs are in
one to one
proportion for reasons of simplicity.  In the locust antennal lobe
the
ratio is more like 3/1 \cite{LL96}. The exact ratios are not
important.

\item The PN/LN pairs are not reciprocally connected. An excited
PN does not directly excite the LN which inhibits it. This is
supported by antennal lobe data \cite{LL96}, \cite{ML96}.

\item Excitatory connections between PNs can be random,
sparse dense, or ``all to all'' without altering the basic properties.

\item Each ensemble is part of a lattice with internal
inhibitory connections. In the simplified models considered here
all PNs receive the same number of inhibitory inputs.

\item The inhibitory connections within ensembles have 
characteristic
time scales different from those between ensembles. Fast and 
slow
inhibition with different pharmacological properties have been
demonstrated in the locust system \cite{LWD96}, \cite{ML96}.

\item Spiking activity in PNs is very low in the absence of
sensory input.
Upon receiving an input, a PN rises above threshold and one or
more action potentials result.
\end{itemize}

To demonstrate how dynamical competition works, we neglect
many
important points about real biological neural assemblies. These
include the action of the sensory inputs on the LNs  \cite{HS97}
and
the variation of synaptic weights over time, perhaps reflecting
adaptation or short-term plasticity in the network. These
processes
are probably essential for the subsequent steps of neural
processing
including learning and recognition, which we do not address
directly
here.

A simple realization of our proposed architecture is seen
in Figure~1b where the lateral inhibitory connections form
hexagons. We assume that the groups of neurons
excited
by different receptors through a glomerulus are not overlapping
and the inhibition between the groups is cyclic.  Despite its
symmetry each of these networks exhibits rich transient
dynamics
upon stimulation. The critical feature of each network is the
absence of reciprocal inhibition of the PNs or of the families of
PNs that form a closed network. Such a connection scheme
could
turn off the excitatory output from one glomerulus and turn on
another. It is crucial to realize that these connectivity schemes
represent {\bf {functional}}, and not simply {\bf {anatomical}}
connections. In other words, many anatomical connectivity
schemes
can be equated to this functional matrix.

The simplified model presented in Figure~1b with high symmetry
and no
random connections is still quite complex. To allow analysis we
further simplify the model. We take all neurons with the same
``color'' within the same ensemble to be tightly synchronized,
and we
treat them as a single neuron. The circuit is then reduced to the
cyclic model in Figure~1c consisting of nine groups of neurons.

We have investigated several implementations of this simplified
model:
(1) ``averaged neurons'' with smoothed voltage responses to a
stimulus, and (2) two dimensional FitzHugh-Nagumo oscillators
with
periodic spiking. We will first describe an ``averaged'' model to
provide some insight into the underlying dynamics of this circuit.
We
will then examine  networks constructed from FitzHugh-Nagumo
neurons.

\subsection*{Averaged Dynamics}

We begin by considering one triplet of coupled neurons from the
nine
neuron model seen in Figure~1c. We characterize each neuron
by its
``activity'' $Y_i(t)\geq0$. The interaction among neurons is
quadratic in 
$Y_i(t)$ reflecting simplified inhibitory synaptic connections
acting 
without the usual threshold and time delay. These $Y_i(t)$ are
taken to satisfy dynamical equations of the form
\be
\frac{ d Y_i(t)}{dt} = Y_i(t) \biggl\{\sigma(E_i(t) + S_i(t))
- \sum_{k=1}^3\rho_{ik} Y_k(t) \biggr\},
\ee
where $\sigma(x)$ is the threshold function
\be
\sigma(x) = 1 - \frac{2}{1 + e^{10(x-0.4)}},
\ee
and
\be
E_i(t) = g_e\sum_{k \ne i}Y_k(t),
\ee
$g_{e}$ is the strength of excitatory connection, $S_i(t)$ is the
input sensory stimulus arriving at neuron $I$, $E_i(t)$ is the
excitatory input from other neurons in the ensemble or triplet,
and
the $\rho_{ik}$ are the strengths of inhibitory synapses within
the same ensemble.

When there is no input, $S_{I}=0$, the threshold function is
$\sigma(x)\approx-1$, each activity is strongly damped and
all $Y_i \to 0$. When an input activates neuron I,  $\sigma (x)
\approx +1$, the dynamics of that neuron near
(0,0,0)---the rest state---becomes unstable and its activity grows
exponentially rapidly towards one of the unstable fixed points
(1,0,0), (0,1,0), or (0,0,1). Because these fixed points are also
unstable, the trajectory approaches one of them, for example,
(0,0,1), and never quite reaches it before the instability of
that fixed point drives it towards another, for example (0,1,0).
The
activity $Y_{3}\approx1$ enters the equation for both $Y_{1}$
and
$Y_{2}$ through $E_{1}$ and $E_2$ and further destabilizes
motion near the
other fixed points. The trajectory, determined in detail by the
stimulus pattern $(S_{1},S_{2},S_{3})$, continues to traverse the
heteroclinic region until the stimulus is removed at which time
the system returns to the rest state.

An input pattern (for example, a spike train from afferents
integrated
to provide a pulse of some longer duration) produces a well
determined, reproducible pattern of firing both in space (neuron
identity) and time. Different input patterns will give rise to
different sequences of waveforms in each individual neuron.

In Figure~2a we show the sequence of waveforms in one of
these
triplets arising from the input choice $S_i = (0.721,0.089,0.737)$
over a fixed time interval (zero otherwise). Its phase space
portrait
$(Y_1(t),Y_2(t),Y_3(t))$ is displayed in Figure~2b. One can see
the
elements of sequential firing observed experimentally as well as
the
basic dynamical idea which is a trajectory transformation along
the
``ribs'' of the heteroclinic connections between the unstable fixed
points of our system. If we change the stimulus to $S_i
= (0.189,0.037,0.342)$ representing a different ``odor'', the
pattern changes markedly as also observed experimentally
(Figure~2c,d).

The next level of complexity introduces a network of three
ensembles
of such triplets interconnected by inhibition. This structure
resembles that within a triplet but we have increased the spatial
scale on which the dynamics operates.  We now have the
following
equations for the activity of the $I^{th}$ neuron in the
$\alpha^{th}$
triplet $Y^{\alpha}_I(t)$
\be
\frac{ d Y^{\alpha}_I(t)}{dt} = Y^{\alpha}_I(t) \biggl\{
\sigma(E^{\alpha}_I(t) + G^{\alpha}_I(t)+ S^{\alpha}_I(t))
- \sum_{k=1}^3\rho_{ik} Y^{\alpha}_k(t) - g_i\sum_kY_k^{
(\alpha+1)\bmod 3}
\biggr\},
\ee
where $E^{\alpha}_I(t) = g_e\sum_{k \ne i}Y^{\alpha}_k(t)$ is the
PN
excitatory input from other neurons within the triplet,
$G^{\alpha}_I(t)$ are excitatory inputs from the other triplets to
element $Y^{\alpha}_I(t)$, and $g_i$ is the strength of
inter-ensemble inhibition. $S^{\alpha}_I(t)$ is the externally
imposed sensory input to neural element $I$ found within neural
triplet $\alpha$.

This network of triplets with constituent averaged neurons
evolves as
described earlier. No single ensemble can remain excited as the
stable
state of the dynamics. Each one is active only transiently,
intermittently or possibly not at all, depending in detail on the
input configuration (see Figure~3). Each input evokes, and is
therefore represented by, a specific trajectory in the state space
of
the system as a whole.

\subsection*{Spiking Neurons}

We next add spiking behavior of the simplest sort to these
``averaged'' neurons by using a FitzHugh-Nagumo model for
each PN in
the network. This provides a limit cycle at each neural node.
While
this addition falls short of a conductance-based description, it
captures another essential feature of the neural dynamics.

The basic triplet of the dynamically
competitive network is now described by a membrane voltage
$U_i(t);
\, I= 1,2,3$, a slow current $I_i(t)$ and a postsynaptic
variable $v_i(t)$ satisfying
\bea
\epsilon\frac{d U_i(t)}{dt} &=& f(U_i(t)) - I_i(t) -\gamma_1(U_i(t) -
\tilde{U}_{I})v_{I-1}(t) +S_i(t), \nonumber \\
\frac{d I_i(t)}{dt} &=& U_i(t) - b I_i(t) +a , \nonumber \\
\alpha_1\frac{d v_i(t)}{dt} &=& \sigma(U_i(t)) - v_i(t) ,
\eea
in which the three FitzHugh-Nagumo neurons are connected by
the
inhibitory coupling $-\gamma_1(U_i(t) - \tilde{U}_{I})v_{I-1}(t) $.
$v_0(t)
= v_3(t)$, $\tilde{U}_{I}=\min(U_{I}(t))\simeq-1.5$, $f(U)= U
- U^3/3$ and $\sigma(U)$ is a step function.  When a stimulus
$S_{I}\geq0.4$ is provided to this triplet, one or more of the
$v_i$
is driven into activity by the increase in membrane voltages. This
in
turn drives the network, via the inhibitory couplings from
neighboring
neurons, along separatrix-like orbits. Our results are shown in
Figure~4. One can clearly see the dynamical competition among
unstable
regions of the state space. Each unstable region is now
characterized
by nearly periodic behavior associated with the limit cycle or
periodic spiking activity.

The final embodiment of our dynamical competition network
combines
three of these spiking triplets in a fashion similar to that for the
averaged neurons above. We associate a fast membrane
potential
$U^{\alpha}_I(t)$ with the $I^{th}$ neuron in the $\alpha^{th}$
ensemble along with a slow current $I^{\alpha}_I(t)$ and two
postsynaptic variables $v^{\alpha}_I(t)$ and $w^{\alpha}_I(t)$
corresponding to fast and slow inhibition respectively.

The dynamics are now specified by equations:
\bea
\epsilon\frac{d U^{\alpha}_I(t)}{dt} &=& f(U^{\alpha}_I(t)) -
I^{\alpha}_I(t) -\gamma_1(U^{\alpha}_I(t) - \tilde{U}^{\alpha}_{ I})
v^{\alpha}_{I-1}(t) \nonumber \\& -&
\gamma_2(U^{\alpha}_I(t) - \tilde{U}^{\alpha}_{I})
\sum_i w^{\alpha-1}_I(t)
+S^{\alpha}_I(t), \nonumber \\
\frac{d I^{\alpha}_I(t)}{dt} &=& U^{\alpha}_I(t) -
b I^{\alpha}_I(t) +a , \nonumber \\
\alpha_1\frac{d v^{\alpha}_I(t)}{dt} &=& \sigma(U^{\alpha}_I(t)) -
v^{\alpha}_I(t) , \nonumber \\
\alpha_2 \frac{d w^{\alpha}_I(t)}{dt} &=& \sigma(U^{\alpha}_I(t)) -
w^{\alpha}_I(t).
\eea
$w^0_i(t) = w^3_i(t)$ and $v_0^{\alpha}(t) = v_3^{\alpha}(t)$.

In Figure~5a we show the time courses for a network composed
of
three triplets of FitzHugh-Nagumo neurons connected as
displayed in
Figure~1c. This is now a set of nine membrane voltages in
which the
evolving firing patterns associated with dynamical competition
are
clearly visible. The global spiking activity resembles the global
spatio-temporal activity observed in our experiments in the
antennal lobe. The time series of two simultaneously recorded
PNs are presented in
Figure~5b.

\section*{Dynamical Competition in the Presence of Error and
Noise}

To be useful our dynamical model must exhibit both the reliability
and
the noise tolerance observed in the biological system. Since any
network is subject to internal and external fluctuations, our model
must produce very similar orbits when noise is present in the
network
and the stimulus. These features are part of the dynamics of our
networks.

\subsection*{Insensitivity of Patterns to Initial Conditions}

Using our ``averaged neurons'' we investigated the critical
question
of reproducibility of the spatio-temporal patterns when the initial
conditions for the network were varied. We selected initial
conditions
uniformly distributed within a sphere of radius R around the
origin of
$Y^{\alpha}_l$ space. This captures the idea that the network
can be
in a set of states near the ``rest state'' when the stimulus arrives.
For each R we calculated the orbits starting in this sphere, and
then
we evaluated the cross-correlation function $C_{\tau}(R)$
between
orbits at $R
> 0$, representing uncertain initial conditions, and
orbits at $R = 0$, representing well specified initial conditions.
\be
C_{\tau}(R) = \frac{\langle (Y^{\alpha}_l(t) -
<Y^{\alpha}_l(t)>_t)|_{R=0}
(Y^{\alpha}_l(t+\tau) - <Y^{\alpha}_l(t+\tau)>_t)|_{R}
\rangle_{t,\alpha,l}}{
(\mbox{Variance at R=0}) \, (\mbox{Variance at R}\ne 0)}.
\ee
The averages were made looking at a window of time t thus
allowing us to determine quantitatively how sections of the whole
time series ``stay together'' for this time. Using the whole time
period of 3000 to 5000 time units during which the stimulus was
presented, we are thus able to examine many windows of
potential coherence. We
evaluated $C_{\tau}(R)$ for $t = 200$ and $t = 400$ time units.
The orbits for $R \ge 0$
were time-delayed relative to one another by an amount $\tau$
selected
to {\bf minimize} cross-correlation. This emphasizes the worst
possible case of overlap between orbits starting from different
initial conditions.  Selecting the initial conditions uniformly in a
finite sphere around the origin in $Y^{\alpha}_l$  space gives
added
weight to larger changes in initial conditions than would a
Gaussian
distribution, also contributing to minimizing cross-correlation.
Finally these averages for each $Y^{\alpha}_l$ were averaged
over all
neurons in the network.

In Figure~6 we plot $C_{\tau}(R)$
for all inhibitory couplings $g_i = 0.5$ over an
averaging window of 200 time steps and
for all $g_i = 0.1$ over averaging windows of 200 and 
400 time units. 
When the inhibition responsible for trajectories along the
heteroclinic orbits is strong ($g_i= 0.5$), $C_{\tau}(R)$ drops
from a
maximum at $R=0$ to a plateau of about 0.84 for $R \geq 0.05$.
This
indicates an 84\% reproducibility of the trajectories of our
system,
namely a remarkable insensitivity to the initial state of the system
at the time of stimulation.

With weaker inhibitory coupling ($g_i = 0.1$), the
cross-correlation
drops only to 0.75, indicating the still enormously high reliability
of
spatio-temporal trajectories in $Y^{\alpha}_l$ space. This small
dependence 
on or ``forgetting'' of the initial conditions relies on the
exponentially
fast departure from rest as the neural assembly is stimulated.
The
precise, quantitative rate of that exponential departure depends
directly on $g_i$.
In the last curve we show $C_{\tau}(R)$ calculated over
400 time units. Because of
intrinsic instabilities in the nonlinear system and the spread of
initial conditions this value is further reduced to about 0.5. 

\subsection*{Robustness of Patterns to Noisy Stimuli}

We now examine the performance of our competitive networks
in the presence of internal and stimulus noise, using
spiking FitzHugh-Nagumo neurons at each of the nine nodes in
the 
network.  Noise was added to each
stimulus so that
\be
S^{\alpha}_I(t) =  0.3 + s^{\alpha}_I g(t) + \eta^{\alpha}_I + 
\delta^{\alpha}_{I}, 
\ee
where  the $ s^{\alpha}_I$ were chosen in the range 0.1 to 0.5; 
the 
function $g(t)$ the simulates time-dependence of stimulus: it
smoothly 
rises from $0$ to $1$ and then falls to $0.1$ during the
presentation 
of the stimulus. The noise terms $\eta^{\alpha}_I$ were uniformly 
distributed variables with maximum values ranging from 1\% to
20\% of 
the stimulus $s^{\alpha}_I$. The terms $\delta^{\alpha}_{I}$ 
with $|\delta^{\alpha}_{I}|=\delta \in [0.0,0.1]$ represent the 
``offset'' of the stimulus. 
Noise with a maximum level close to the 
firing threshold was added to each dynamical variable, the
$U^{\alpha}_I, 
I^{\alpha}_I, $ etc of the Fitzhugh-Nagumo equations. 

Figures 7a and 7b plot spatio-temporal patterns obtained with
respectively
1\% and 15\%stimulus noise $\eta^{\alpha}_{I}$.
at $\delta=0$. In Figure 7a 
the noise level in the system was below threshold and no spikes 
were generated in the absence of a stimulus. In Figure 7b, we
raised 
the noise level above threshold, and spikes were generated
randomly 
even in the absence of a stimulus. 

Can the dynamical competitive network reproduce the patterns 
associated with a particular stimulus even in the presence of
noise?
We estimated the topological similarity of 
patterns produced with different levels of noise by calculating the
overlap 
$G_{I}^{\alpha}(\delta, \tau, \eta)$ of their spike envelopes 
with and without stimulus noise $\eta_{I}^{\alpha}=0$, for each
neuron.
This overlap is a function of the 
variance in the stimulus $\delta$ and the time scale of the
dynamics 
$\tau$. The time scale is important because 
noise keeps the orbit from approaching the heteroclinic orbits
and 
changes the response time scale for spiking. 

To characterize reproducibility over the whole network, we 
evaluated the product of these 
functions $G(\delta,\tau,\eta)=\prod_{I,\alpha} (G_{I}^{\alpha} ( 
\delta, \tau, \eta))$ over all the neurons. The 
normalized functions $G/G_{max}$
for noise levels $\eta=0.01$ and $\eta=0.15$ are shown in
Figures 7c 
and 7d. Even in the presence of substantial noise, 
the peak in the overlap function is very sharp and appears near
the 
original stimulus at $\delta=0$, 
indicating that the 
competitive network can reliably ``recognize'' the stimulus in the
presence of 
noise.  

\section*{Discussion}

We have proposed an innovative class of neural network models
whose
dynamics reproduce the rich spatiotemporal features observed in
insect
olfactory networks.  Our goal was to determine whether this
seemingly complex form of information coding, whose biological
relevance has been shown experimentally \cite{WL96},
\cite{SBSL97},
\cite{MBL98} presents clear and possibly general advantages for
pattern
classification, memory and recognition. We found that
stimulus-dependent activity of this type, though indeed spatially
and
temporally complex, has underlying dynamical order and stability. 
The
physical principles of this mode of representation rely on
transient
orbits moving between unstable regions of state space (fixed
points or
limit cycles in our models). Each orbit is defined by an input and
has
rich structure only as long as the input lasts.  This is in strong
contrast with the behavior of familiar competitive network
structures.
In the pioneering work of Hopfield, Cohen and Grossberg
\cite{H82},
\cite{CG83} for example, input patterns play the role of initial
conditions which lead the network to store information or
represent
stimuli as attractors. These attractors might be fixed points, limit
cycles or even strange attractors \cite{F87}.

Besides this attractor based storage, we propose that the brain
uses ``winnerless'' networks, in which dynamical competition
between
groups of interconnected neurons produces stimulus-specific
orbits in
their phase space.  In these neural networks, the selection of the
system orbit is strongly influenced by network parameters and by
nonstationary external inputs as found in a realistic environment.
As
we have shown here one essential feature of this dynamical
mode of
representation is that it is remarkably resistant to noise and to
wide
variations in initial conditions. This conclusion holds even though
the trajectories in phase space are not directed towards
traditional
attractors while the input is present.  The fact that
stimulus-specific
trajectories depend {\it continuously} on the input is also a crucial
feature of these networks.  This continuous dependence means
that
stationary states of the system, as found naturally {\it between}
activations by sensory inputs, do not represent stimulus memory
traces.  In other words, the network resets itself ``for free'' as
soon as stimulation ceases.  A fundamental advantage of this
mode of representation is that memory capacity is enlarged. 
Assuming
that input memories incorporate these dynamical features,
interferences between stored patterns will be reduced-each
pattern
``traverses'' the network's phase space along a different
skeleton-and
the probability of a ``superposition catastrophe'', familiar to
classical attractor networks  \cite{VDM81}, will be reduced.

The idea that the transient, stimulus-dependent
(non-autonomous)
behavior of a dynamical system can be used to encode
information
is, to our knowledge, not found directly in the dynamical systems
literature. Nothing that we know suggests that such behavior has
singular or unacceptable mathematical properties. Fortunately,
the
systems we have used to explore and illustrate our ideas are
simple,
which should allow mathematical analyses in the future.  Our
ideas can
be linked to a class of control algorithms whose key features rest
on
selecting a time-dependent input to stabilize an unstable periodic
orbit of the autonomous dynamical system (see \cite{OGY90}). 
Our
goal, however, is fundamentally different:  we (or nature)
exploit(s)
the rich properties of the stimulus-driven system instead of
stabilizing the autonomous (resting) system, whose state is
stable.

It is also interesting to note that the ``non-potential'' behavior of
the trajectories which traverse state-space close to heteroclinic
structures has precursors in the dynamical theory of pattern
formation.  Busse and Heikes (1980) \cite{BH80} proposed a
model for
the description of convection in rotating fluids which consisted of
three variables representing three sets of fluid rolls oriented at
$\frac{2\pi}{3}$ relative to each other.  Their model has no stable
fixed point, as in our simplified triplet model with stimulus, and
evolves along the heteroclinic orbits connecting the three
unstable
fixed points.

Our dynamical encoding scheme raises a number of important
biological
questions. For example, if the brain uses such strategies to
represent
sensory stimuli \cite{PCPM97}, \cite{M86}, \cite{K74}, how are
such
dynamical patterns decoded? Although this is ultimately an
empirical
question, it is useful to examine the implications of our proposal.
First, we note that temporal decoding algorithms may not pose
fundamental theoretical problems (\cite{BM95}, \cite{BM98},
\cite{M89}). Second, physiological and behavioral experiments
carried
out with the insect olfactory system (and which motivated this
study)
already indicate
that time is a relevant parameter for stimulus representation and
decoding (\cite{LN94}, \cite{WL96}, \cite{SBSL97}, \cite{MBL98}).
Third, note that even if a decoder of spatio-temporal patterns
had a
long time constant, i.e., compressed a spatio-temporal
representation into a spatial  one, our representation would have
still gained in precision, because of the intrinsic stability of the
dynamical encoding\footnote{It is important to note that 
such a mode of representation will work best if
the time scale of the stimulus is slower than those of the
networks
representing it}.

The properties of these simplified model networks may
also emerge from more realistic circuits having the same general
character of dominance by unidirectional inhibitory connections.
In
particular, if the system is large, it is not necessary to demand
that
the network architecture be cyclic. 
A network with sparse, random
connections will exhibit the same essential features as our
simple,
example networks. Indeed, we anticipate that more complex and
more
realistic networks of neurons and synapses will make dynamical
competition behavior even more advantageous and natural for
representation, and possibly learning, storage and recognition of
sensory inputs. The fundamental scheme we propose is not
linked to the
nature of the stimulus, though it was clearly suggested by our
analysis of odor inputs, nor is it linked to the size of the brain
processing the stimulus. The strategy we have described may
thus apply
to many other brain circuits in insects and other animals,
including
mammals and constitute a fundamental feature of sensory
networks in
the brain.

\section*{Acknowledgments}
We thank members of INLS and the Laurent laboratory for
extensive
discussions, and Christof Koch for helpful comments on a draft
of this paper.
This work was supported in part by the U.S. Department of
Energy,
Office of Basic Energy Sciences, Division of Engineering and
Geosciences, under grant DE-FG03-90ER14138, in part by
ORD, and by
NIDCD (Laurent).

\clearpage
\section*{Figure Captions}
\begin{itemize}
\item [{\bf Figure 1}] (1a.) The general scheme of dynamically
competitive neural models. This is a single triplet of neurons in
$(Y_1,Y_2,Y_3)$ space. The unstable fixed points (saddles) at
(1,0,0),
(0,1,0) and (0,0,1) are connected by three separatrices, the
``ribs''
in our description. When the network receives a stimulus, an
orbit in
this space visits the neighborhood of these ``ribs'' as it move
from
fixed point to fixed point. (1b) An example of a neural network
architecture with ``no winner'' competition. We portray a set of
three
groups of PN/LN neurons each organized into a hexagonal
lattice. There
is unidirectional inhibition between the lattices causing activation
of this network to proceed from lattice to lattice in a ``no winner''
fashion. (1c) Reduction of three lattices of neurons to  simplified
ensembles of three PNs (red circles) and three LNs (blue circles)
coupled with unidirectional inhibition shown by blue arrows and
excitatory connections shown by red arrows.

\item [{\bf Figure 2}] (2a) Time series of a single triplet of
averaged neurons, Equations (1)--(3), with inputs $S_i =
(0.721,0.089,0.737)$. (2b) The state space orbit
$(Y_1(t),Y_2(t),Y_3(t))$ for the times series shown in Figure 2a.
(2c)
Time series of a single triplet of averaged neurons, Equations
(1)--(3), with inputs $S_i = (0.189,0.037089,0.342)$ (another
``odor''). (2d) The state space orbit $(Y_1(t),Y_2(t),Y_3(t))$ for
the
times series shown in Figure 2c. We chose $\rho_{11} =
\rho_{22} =
\rho_{33}
=1$, $\rho_{12} = \rho_{23} = \rho_{31} = 5$ and $\rho_{21} =
\rho_{32} = \rho_{13} = 0.2$ along with $g_e = 4$.

\item [{\bf Figure 3}] The time series of neurons 1,~4,~8 from
three
different triplets of averaged neurons from equation~(4) with
$g_{I}=0.1$, $S_{1}^{1}=1.0$, and the same parameters
$\rho_{ij}$ as
in figure~(2). In spite of the stimulus was applied only to one
neuron, all neurons are firing.

\item [{\bf Figure 4}] Three dimensional phase portrait of the
dynamical competition for a network of three FitzHugh-Nagumo
neurons,
Equation (5). Note the ``winnerless'' movement between the
spiking
behavior (unstable limit cycles) of the neurons as the orbits
traverse
the network upon stimulation. Here we have taken $\epsilon =
0.08, a =
0.7, b = 0.8, \alpha_1 = 10.1, \gamma_1 = 2.0, S_i =
(0.39,0.4,0.399).$ This figure uses state space coordinates which
are
linear combinations of the FitzHugh-Nagumo variables: $\xi_1 =
15 v_1
+ I_1 + U_3, \xi_2 = 15 v_2 + I_2 + U_1,$ and $\xi_3 = 15 v_3 +
I_3 +
U_2$.

\item [{\bf Figure 5a}] Time series of nine noisy
FitzHugh-Nagumo
neurons organized into three families connected by unidirectional
inhibition, Equation (6). We chose $\epsilon = 0.08, a = 0.7, b =
0.8,
\alpha_1 = 5.1, \alpha_2 = 50.1, \gamma_1 = \gamma_2 = 2.0,
S_i =
(0.5,0.5,0.6,0.5,0.5,0.5,0.5,0.5,0.9)).$ The stimulus strength
$(S_{I}-0.3)$ is shown in green bars on the right of each time
course.
The stimuli were on for times $1000 \le t \le 3000$, and
independent,
identically distributed noise uniform in $[0.0, 0.015]$ was added
to
each stimulus. The ``magnifying glass'' shows individual spikes
generated by one neuron.

\item [{\bf Figure 5b}]  Simultaneous intracellular recordings from
two PNs in the locust antennal lobe. Here the time is in ms; the
stimulus was applied for time [1000,2000].
%and extracellular local field
%   potential recording from their target area, the mushroom
body.
Note the complex response patterns of these two neurons,
comprising
fast, transient oscillatory activity (also seen in the local field
potential) as well as slower and neuron-specific modulation of
firing
rate, in response to the presentation of the odor heptanone.
Methods
as in \cite{LWD96}.

\item [{\bf Figure 6}] The cross correlation $C_{\tau}(R)$ for the
model in Equation (4) as a function of R, the radius within which
initial conditions for the dynamical competition network were
uniformly distributed. The cross correlation functions were
normalized
to $C_{\tau}(R = 0)= 1$.

\item [{\bf Figure 7}] Response of orbits to stimulus noise.
Spiking activity of the nine
neuron model of dynamical competition using FitzHugh-Nagumo
neurons at 
each node (Equation (6)) with 1\% and 15\% noise (a, b). We used
the stimulus $s_{1}^{1}=0.1$, 
$s_{1}^{2}=0.1$, $s_{1}^{3}=0.2$, $s_{2}^{1}=0.2$,
$s_{2}^{2}=0.3$, 
$s_{2}^{3}=0.2$, $s_{3}^{1}=0.2$, $s_{3}^{2}=0.1$, and
$s_{3}^{3}=0.1$, 
applied during $1000 \le t \le 3000$. It was slowly increased
from 0 
to 1 over $1000 \le t \le 1500$ and then slowly decreased from 1
to 
0.1 during $1500 \le t \le 3000$ as shown in the top traces.
The noise level was $\eta_{I}^{\alpha} = 0.01$ (a)
and $\eta_{I}^{\alpha}= 0.15$ (b). The stimulus offset 
was $\delta=0$. c,d Overlap between 
the neural firing patterns with and without noise.
 $\eta_{I}^{\alpha} = 0.01$ in (c)
and $\eta_{I}^{\alpha} = 0.15$ (d). The stimulus 
``offset'' was $\delta_{1}^{1}=\delta$, $\delta_{1}^{2} 
=-\delta$, $\delta_{1}^{3}=-\delta$, $\delta_{2}^{1}=\delta$, 
$\delta_{2}^{2} = -\delta$, $\delta_{2}^{3}=\delta$, $\delta_{3}^{1} 
=\delta$, $\delta_{3}^{2} =-\delta$, $\delta_{3}^{3}=\delta$. 

\end{itemize}

\end{document}